\documentclass[aps,pre,nofootinbib,twocolumn,showpacs,preprintnumbers,amsmath,amssymb,floatfix,superscriptaddress,10pt]{revtex4}
\usepackage{amsmath}
\usepackage{amsfonts}
\usepackage{dcolumn}                    	     % Align table columns on decimal point
\usepackage{amssymb}
\usepackage{graphicx,graphics,wrapfig,rotating}     	% Include figure files
\usepackage[english]{babel}	
\usepackage{bm,fancybox}                	     % bold math

\begin{document}

\title{Transient features of quantum open maps}

\author{Leonardo Ermann} 
\email[Email address: ]{ermann@tandar.cnea.gov.ar}
\affiliation{Departamento de F\'\i sica Te\'orica, GIyA, Comisi\'on Nacional de Energ\'ia At\'omica, Buenos Aires, Argentina}

\author{Gabriel G. Carlo} 
\affiliation{Departamento de F\'\i sica Te\'orica, GIyA, Comisi\'on Nacional de Energ\'ia At\'omica, Buenos Aires, Argentina}

\author{Juan M. Pedrosa}
\affiliation{Departamento de F\'\i sica Te\'orica, GIyA, Comisi\'on Nacional de Energ\'ia At\'omica, Buenos Aires, Argentina}
\affiliation{Escuela de Ciencia y Tecnoloig\'ia, Universidad Nacional de San Mart\'in, San Mart\'in, Buenos Aires, Argentina}

\author{Marcos Saraceno} 
\affiliation{Departamento de F\'\i sica Te\'orica, GIyA, Comisi\'on Nacional de Energ\'ia At\'omica, Buenos Aires, Argentina}
\affiliation{Escuela de Ciencia y Tecnoloig\'ia, Universidad Nacional de San Mart\'in, San Mart\'in, Buenos Aires, Argentina}

\date{\today}

%05.45.Mt Quantum chaos; semiclassical methods

\begin{abstract}

We study families of open chaotic maps that classically share the same asymptotic properties -- forward 
and backwards trapped sets, repeller dimensions, escape rate -- but differ in their short time behavior. 
When these maps are quantized we find that the fine details of the distribution of resonances and the 
corresponding eigenfunctions are sensitive to the initial shape and size of the openings. We study 
phase space localization of the resonances with respect to the repeller and find strong delocalization 
effects when the area of the openings is smaller than $\hbar$.

\end{abstract}

\maketitle

\section{Introduction}

The quantum treatment of scattering situations where the trapped set has a complex fractal structure 
has recently been the source of much attention, particularly with respect to the 
confirmation of the fractal Weyl law \cite{Lu,Schomerus,Nonnenmacher2005} hypothesis concerning the 
global distribution of long lived resonances and their localization properties on the classical 
repeller \cite{Keating,Wiersig,Carlo}. The use of quantum maps to model these systems has been 
of paramount importance, allowing extensive numerical, and sometimes, analytical results. 

A classical ``closed'' map $\mathcal{B}$ is 
represented as a diffeomorphism on a symplectic manifold $\Sigma$ \cite{Nonnenmacher2011}.
Its open version $\tilde{\mathcal{B}}$ can be defined restricting the phase 
space to $V$ (a subset of $\Sigma$), which means that trajectories outside 
of these region escape. The associated classical repeller $\mathcal{K}$ 
is defined as the set of trajectories which remain on $V$ both in the past and the 
future for infinite times (i.e. the intersection of the backwards and forward trapped set). 
If the closed map 
$\mathcal{B}$ is chaotic, the dynamics of $\tilde{\mathcal{B}}$ on 
the repeller is hyperbolic and complex, and the resulting $\mathcal{K}$ 
has a fractal structure with non-integer dimension $d$. Also, 
an initially uniform probability distribution in phase space decays exponentially 
with the classical escape rate $\gamma$.
These are asymptotic features of the map appearing as $T \rightarrow \infty$, 
that are associated to the global distribution of resonances and how their 
number scales with $\hbar$.
Note that given an open map we have a unique associated repeller, 
but the converse is not true. A given repeller can be related to a large set of open maps 
(e.g. for the same $\mathcal{B}$, the repeller given by $V$ 
and $V''=\tilde{\mathcal{B}}(V)$ are the same). 

When quantized, the evolution of these systems is non-unitary, there is a probability loss reflecting 
the classical loss of trajectories. The corresponding operators have right and left 
decaying nonorthogonal eigenfunctions with complex
eigenvalues $\lambda_j$, also referred to as resonances. They fall inside the
unit circle in the complex plane ($|\lambda_j|^2=\exp{(-\Gamma_j)} \leq 1$). 
The exponent $\Gamma_j \geq 0$ can be interpreted as the rate at which probability is lost and 
thus is called the (quantum) decay rate. The long lived resonances are those with the 
lowest decay rates and are directly associated to the classical repeller $\mathcal{K}$. 
Their $\Gamma_j$ are related to the classical escape 
rate $\gamma$ in the generic case (see however \cite{Pedrosa}).

In this paper we study classical maps having the same repeller and escape rate but 
different openings, and their corresponding quantum counterparts, by means of the 
paradigmatic tri-baker map. How these differences affect the behavior of the resonances is a 
fundamental question. Moreover, this is a crucial 
step in order to understand and evaluate the performance of the semiclassical 
theory attempting to describe open maps \cite{SemiclassicalTh}. Out of the many ways to open a map which lead to the 
same repeller we have found that two of them are the most important for this purpose, 
giving rise to two families of maps. 
Keeping the area of the opening constant (\emph{shift} family) or letting it to increase 
(\emph{intersection} family) defines them and is the source of important dynamical differences 
during the first time steps (though the asymptotic decay rate is the same for both of them). 
In the first case, we have found that by just varying 
the shape of the opening, the behavior of the long lived portion of the spectrum 
changes slightly, in a non trivial fashion. 
In the second case, the quantum spectrum changes as a whole.
Finally, the \emph{shift} family allows to identify a quantum phase transition in the localization of 
the resonances, for openings having areas smaller than $\hbar$.

In the following we describe the organization of this paper. In Section \ref{sec:Openmaps} we define the 
family of maps that we study and their quantizations. In Section \ref{sec:Results} we analyze the properties of 
the spectrum and the corresponding set of eigenfunctions by means of different measures and a phase space distribution 
that is specially useful for open maps \cite{ermannprl}. Finally, in Section \ref{sec:Conclusions} 
we point out our concluding remarks.

\section{Open maps}
\label{sec:Openmaps}

Given a repeller $\mathcal{K}$ there are many open 
maps that have it as the invariant set. 
We will take $V_1$, an opening in phase space, as the generator of all these open 
maps with the same repeller. 
For simplicity, in our example (which follows) it will be taken symmetric in $q$ and $p$ 
(invariant under $q\leftrightarrows p$ exchange). Other choices that do not respect 
this symmetry (or equivalent ones in other cases) lead to more differences in the 
resonances whose study is out of the scope of the present paper \cite{future}. 
The next step in the construction consists of performing the evolution of $V_1$ 
under $\mathcal{B}^{t}$, that will be called $V_{t}\equiv \mathcal{B}^{t-1}(V_{1})$.
Finally, simple combinations of $V_t$ for positive and negative 
times allow to create the open maps, which can be grouped into map families according 
to their properties, all of them sharing the same repeller $\mathcal{K}$.
Moreover, iteration of open maps $\tilde{\mathcal{B}^t}$ can be used alternatively 
to build new maps corresponding to this invariant set.

We will focus into two families of maps that are essentially given by preserving the area of 
the opening for all of its members in one hand or not preserving it on the other.
These families are the most interesting since they allow to understand the dependence of the 
quantum spectrum on the shape and on the size of the opening, respectively while keeping the repeller 
and asymptotic decay rate invariant. 
The members of the first family $\tilde{\mathcal{B}}^{s}_k$ with $k=1,2,\ldots$, 
which we call \emph{shift} family, 
are defined as one iteration of the closed map $\mathcal{B}$
followed by the escape of trajectories outside the region $V_{k}$.
In turn, the \emph{intersection} family is defined as
$\tilde{\mathcal{B}}^{i}_k$ with $k=1,2,\ldots$, where the allowed region
this time is given by the intersection of regions until time $k$,
$V_{k}\cap V_{k-1}\cap\ldots \cap V_{2}\cap V_{1}$.
For this family, the allowed region decreases for larger $k$ members.
Of course, for nontrivial $\mathcal{B}$ and $V_1$ all family members are different, though 
the first member of both families are exactly the same by definition 
($\tilde{\mathcal{B}}^{s}_1\equiv\tilde{\mathcal{B}}^{i}_1$).

\subsection*{The tri-baker map}

\begin{figure}%[htp]
\begin{center}
\includegraphics[width=0.4\textwidth]{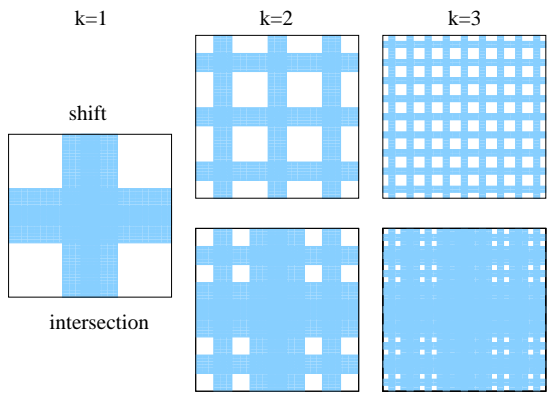} 
\end{center}
\caption{(Color online) Phase space representation of the openings for \emph{shift} and \emph{intersection} families 
with $k\le3$, in top and bottom panels respectively.
The allowed regions are shown in white while the corresponding openings are in light blue (gray).}
\label{fig1}
\end{figure}

We have chosen the paradigmatic tri-baker map as the model for our studies.
This is one of the simplest chaotic maps
which can be easily described by a ternary Bernoulli shift, and where the openings can be done following 
stable and unstable manifold directions. The general open map is defined as the composition of the 
\emph{closed} tri-baker transformation followed by a given opening.
We will define both \emph{shift} and \emph{intersection} families
generated by an initial allowed region $V_1$ having two horizontal strips $p\in[0,1/3)\cup (2/3,1)$.
The openings will be defined to be symmetrical in $q$ and $p$, in such a way as to maintain the 
time reversal symmetry of the closed map as shown in the left panel of Fig.\ref{fig1} ($k=1$).

The tri-baker map in a unit square phase space 
$\mathbb{T}^2\equiv[0,1)\times[0,1)$ is given by
\begin{equation}
 \left( \begin{array}{c} q^\prime \\p^\prime \end{array} \right)=
\mathcal{B}  \left( \begin{array}{c} q \\p \end{array} \right) = 
 \left( \begin{array}{c} 3q-[3q] \\ (p+[3q])/3 \end{array} \right)
\end{equation}
where $[q]$ means the integer part of $q$. 
The map is uniformly hyperbolic with Lyapunov exponent $\lambda=\log3$.
The symbolic notation of the map action is given by 
a Bernoulli shift in ternary representation of $q=0.\epsilon_0 \epsilon_1 
\epsilon_2 \ldots$ and $p=0.\epsilon_{-1} \epsilon_{-2} 
\epsilon_{-3} \ldots$ (given by the corresponding 
trits $\epsilon_i=0,1,2$) as
\begin{equation}\label{eq.shift}
(\mathbf{p}\vert\bf{q})=...\epsilon_{-2}\epsilon_{-1}.\epsilon_{0}\epsilon_{1}...
\xrightarrow{\mathcal{B}}
(\mathbf{p}^\prime\vert\bf{q}^\prime)=...\epsilon_{-2}\epsilon_{-1}\epsilon_{0}.\epsilon_{1}...,
\end{equation}
where the dot is moved one position to the right. Different openings of both \emph{shift} 
and \emph{intersection} families can be straightforwardly 
defined in ternary notation using open trits $\tilde{\epsilon}$ with forbidden value 1 ($\tilde{\epsilon}=0,2$).

\emph{Shift} family members $\mathcal{B}^{s}_k$ have two open trits corresponding to the $k$-th most
significant trit of both position and momentum in ternary representation.
On the other hand, \emph{intersection} family members have the first $k$ open trits of both position and momentum.
In order to illustrate \emph{shift} and \emph{intersection} family members,
 $\mathcal{B}^{s}_3$ and $\mathcal{B}^{i}_3$ are shown in the next equation with open trits highlighted inside boxes 
\begin{eqnarray}
 \mathcal{B}^{s}_3&\longrightarrow& \ldots\epsilon_{-5}\epsilon_{-4}\boxed{\tilde{\epsilon}_{-3}}\epsilon_{-2}\epsilon_{-1}.
\epsilon_{0}\epsilon_{1}\boxed{\tilde{\epsilon}_{2}}\epsilon_{3}\epsilon_{4}\ldots\\
 \mathcal{B}^{i}_3&\longrightarrow& \ldots\epsilon_{-5}\epsilon_{-4}\boxed{\tilde{\epsilon}_{-3}\ \tilde{\epsilon}_{-2}
\tilde{\epsilon}_{-1}.\tilde{\epsilon}_{0}\tilde{\epsilon}_{1}\ \tilde{\epsilon}_{2}}\epsilon_{3}\epsilon_{4}\ldots
\end{eqnarray}
The first 3 members of both families are represented geometrically in Fig.\ref{fig1}.
In this way the classical escape rate can be computed analytically. 
When a closed trit is opened (transformation from $\epsilon_j\rightarrow\tilde{\epsilon}_j$) 
the area of the allowed space is reduced by a factor $2/3$ ($A\rightarrow 2/3 A$).

For the case of the $k$ member of the \emph{shift} family, two trits are opened in the first iterations of the map 
until the first open trit of position ($\epsilon_{k-1}$) reaches the $k$-th most significant trit of momentum. 
From then on the map only opens one trit in each iteration.
Therefore, during the first $2k$ iterations the allowed area decreases by a factor of $(2/3)^2$ in each step, and from then the 
area decreases by a factor of $2/3$. The same reasoning applies to the $k$ member of the \emph{intersection} family, where 
$2k$ trits are opened in the first iteration, and from then only one closed trit is opened at each step.
Therefore the allowed area in the first step decreases suddenly by a factor of $(2/3)^{2k}$, but from the second 
step decreases by $(2/3)$ and so on. Thus the asymptotic escape rate is the same for both families.

% One of the interesting points of these two families is that 
% they do not only converge to the same repeller, but also have the same escape rate (asymptotically). 
% Examples of maps that converge to the same repeller with different escape rates can be obtained as iterations 
% of this kind of maps (e.g. for $\mathcal{B}^{\prime}\equiv(\mathcal{B}^{i}_1)^2$ the escape rate is two times larger).

\subsubsection*{Quantum version}

There are different ways to quantize the tri-baker map \cite{ermannessential}. 
We will follow the 
Balazs-Voros-Saraceno (BVS) quantization, which is done by choosing anti-periodic boundary 
conditions in $q$ and $p$ \cite{bakerbvs}. The 
Hilbert space dimension is given by the integer $D=1/(2\pi\hbar)$
with position and momentum eigenvectors $\vert q_j\rangle$ and $\vert p_j\rangle$ 
(where $j=1,\ldots,D$) connected by the anti-symmetric Fourier transform $G_D$, given by 
\begin{equation}\label{eq.ft}
 \left(G_D\right)_{j^\prime,j}\equiv\langle q_{j^\prime}\vert p_j\rangle=\frac{1}{\sqrt{D}}
e^{-i\frac{2\pi}{D}(j^\prime+\frac{1}{2})(j+\frac{1}{2})}.
\end{equation}
The quantum baker map on a Hilbert space of dimension $D=3M$ can be achieved converting
the most significant qutrit of position in the most significant qutrit of momentum. 
The matrix of the map in mixed representation has the form of three diagonal blocks with a 
finite Fourier transform of size $D/3$ in each one of them. 
In position representation we have
\begin{equation}\label{eq.baker}
B_{pos}=G^\dagger_D B_{mix}=G^\dagger_D \left(\begin{array}{ccc}
 G_{D/3}&0&0\\0&G_{D/3}&0\\0&0&G_{D/3}\\\end{array}\right)
\end{equation}
The map is opened by means of a projector $\Pi$, and in order to respect the $q \leftrightarrows p$ 
symmetry we will use quantum open maps of the form $\tilde{B}_{mix}=\Pi B_{mix}\Pi$ 

The quantum version of the \emph{shift} and \emph{intersection} families can be easily implemented 
for the quantum tri-baker map on $l$ qutrits with $D=3^l$.
The forbidden $1$ in one trit $\tilde{\epsilon}$, can be quantized by the one qutrit 
projector $\pi=I-\vert1\rangle\langle 1\vert=\vert0\rangle\langle 0\vert+\vert2\rangle\langle 2\vert$. 
In this way, the projector applied to the $i$-th qutrit can be written as 
\begin{equation}
\Pi_i=\underbrace{I\otimes \ldots \otimes I}_{i-1}\otimes \pi \otimes 
\underbrace{I\otimes \ldots \otimes I}_{l-i}  
\end{equation}
with $\Pi_i=\Pi_i^\dagger$, $\Pi_i^2=\Pi_i$ and $[\Pi_{i},\Pi_{j}]=0$ where $i,j=1,\ldots,l$.
Following this notation, the \emph{shift} and \emph{intersection} families of open quantum baker 
maps can be straightforwardly defined as
\begin{eqnarray}
 \tilde{B}^{s}_k&=&G^\dagger_D\Pi_{k}B_{mix}\Pi_{k}\\
 \tilde{B}^{i}_k&=&G^\dagger_D
\Pi_{1}^\dagger
\ldots\Pi_{k-1}^\dagger\Pi_{k}^\dagger
B_{mix}\Pi_{k}\Pi_{k-1}\ldots\Pi_{1}
\end{eqnarray}
with $k=1,\ldots,l$, respectively.

A phase space representation of the openings for \emph{shift} and \emph{intersection} families 
with $k\le3$ is shown in Fig. \ref{fig1}. 

\section{Results}
\label{sec:Results}

\subsection{Distribution of resonances}

\begin{figure}
\begin{center}%[htp]
\includegraphics[width=0.46\textwidth]{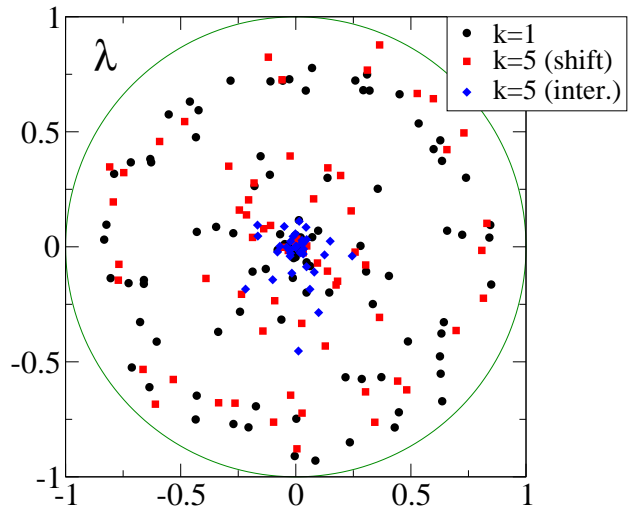} 
\end{center}
\caption{(Color online) Spectrum in the complex plane for three different BVS quantum open baker 
maps with dimension $D=3^5=243$. Black circles represent the eigenvalues for the first member 
of both the \emph{shift} and \emph{intersection} families ($k=1$), 
red squares stand for \emph{shift} family with $k=5$, and blue diamonds for \emph{intersection} with $k=5$.}
\label{fig2}
\end{figure}

The quantum open maps $\tilde{B}$ defined in Sec. \ref{sec:Openmaps} 
are represented by non-normal matrices of dimension $D$ 
($[\tilde{B},\tilde{B}^\dagger]\ne0$). Therefore, their spectrum and 
eigenstates are given by 
\begin{eqnarray}
 \tilde{B}\vert R_j\rangle&=&\lambda_j\vert R_j\rangle\\
 \langle L_j\vert \tilde{B}&=&\lambda_j\langle L_j\vert
\end{eqnarray}
where the complex eigenvalues $\lambda_j$ are inside unit circle $\vert\lambda_j\vert\le1$ 
and $\vert R_j\rangle$ and $\vert L_j\rangle$ are the right and left eigenstates with $j=1,\ldots,D$. 
For a given eigenvalue, the corresponding left and right eigenstates are not equal in general, 
but they obey the orthogonality rule $\langle L_j\vert R_{j^\prime}\rangle\propto \delta_{j,j^\prime}$.
In Fig. \ref{fig2} we show the spectrum corresponding to both families. It can be seen that for 
the first member of each family the results are the same as we expected by definition. But for the 
last members allowed by the dimension $D=3^5=243$, i.e. $k=5$ we find great differences. 
In fact, the one belonging to the \emph{shift} family has a spectrum similar to the previous case while 
the one of the \emph{intersection} family is highly contractive (it shrinks towards zero) due to the 
larger area of the openings.

\begin{figure}
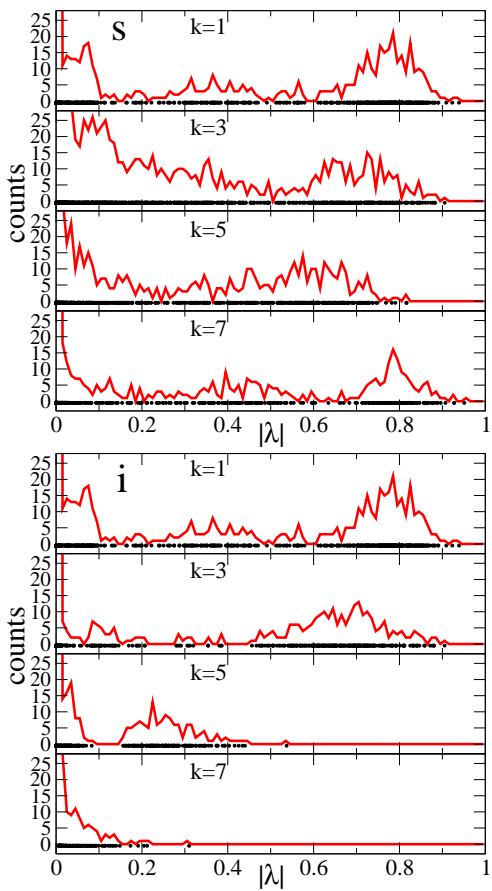
%[htp]
\begin{centering}
\includegraphics[width=0.36\textwidth]{fig3a.eps}
\includegraphics[width=0.36\textwidth]{fig3b.eps} 
\par\end{centering}
\caption{(Color online) Distribution of resonances for \emph{shift} (top panel) and \emph{intersection} 
(bottom panel) quantum open baker maps with dimension $D=3^7=2187$.
Moduli of eigenvalues are represented by black circles for $k=1,3,5,7$ from top to bottom, 
respectively. Histograms of these moduli for 100 intervals from 0 to 1 are shown by means of red lines.}
\label{fig3}
\end{figure}

In order to have a more detailed idea about the shape of the spectrum of both families, 
in Fig. \ref{fig3} we show the histograms corresponding to the distributions of the moduli 
of their eigenvalues for $D=3^7=2187$. 

It can be seen that the \emph{shift} family has a non monotonic behavior with respect to the significance 
of the open trit. The long lived sector of the resonances, clearly visible as a big bell towards the 
rightmost part of the histograms, ``oscillates'' from the right to left and back, while more or less keeping its 
shape. This oscillation shows that, though the classical escape rate is the same for all the maps, the average 
value of the decay rate of the long lived states changes. 

In the case of the \emph{intersection} family there is a completely different behavior. At very short times 
(in fact, from the second time step on) the classical escape rate settles down to its asymptotic value. But the allowed area 
decreases suddenly also. This fact translates into the sudden shrinking of the spectrum as $k$ increases. 
Indeed, the last member of this family allowed by the dimension $D$ has all its eigenvalues 
heavily concentrated around modulus zero. Then, it seems that one has to be aware of short time dynamics at the 
time of counting resonances; i.e., not only the escape rate and fractal dimension counts.

\subsection{Eigenstates and the quantum repeller}

We investigate the morphology of the eigenstates of the open tri-baker map 
families by making use of a recently introduced phase space representation 
suitable for open systems \cite{ermannprl}. The first step is to 
define the symmetrical operators of right and left eigenstates
\begin{equation}\label{eq.hdef}
 \hat{h}_j=\frac{\vert R_j\rangle\langle L_j\vert}{\langle L_j\vert R_j\rangle}
\end{equation}
which are related to the eigenvalue $\lambda_j$ and are independent from eigenstate normalization. 
These operators have the advantage that they clearly show 
the repeller structure underlying in the eigenstates. 
They are equivalent to projectors on the eigenstates of unitary evolution 
operators. To have a better idea of how the set of 
long lived resonances distribute on the classical repeller we also study the generalized 
operators onto this set taken as the sum of the 
first $j$ of the $\hat{h}_{j'}$ projectors, ordered by decreasing modulus of the corresponding eigenvalues  
($\vert\lambda_j\vert\geqslant\vert\lambda_{j^\prime}\vert$ with $j\leq j^\prime$)
\begin{equation}
\hat{Q}_j\equiv\sum_{j^\prime=1}^j\hat{h}_{j^\prime}.
\end{equation}
From the definition of $\hat{h}_j$ in Eq. \ref{eq.hdef} we have that $tr(\hat{Q}_j)=j$ 
and $\hat{Q}_{j}^2=\hat{Q}_{j}$, but $\hat{Q}_{j}^\dagger\neq\hat{Q}_{j}$.  

\begin{figure}%[htp]
\begin{center}
\includegraphics[width=0.48\textwidth]{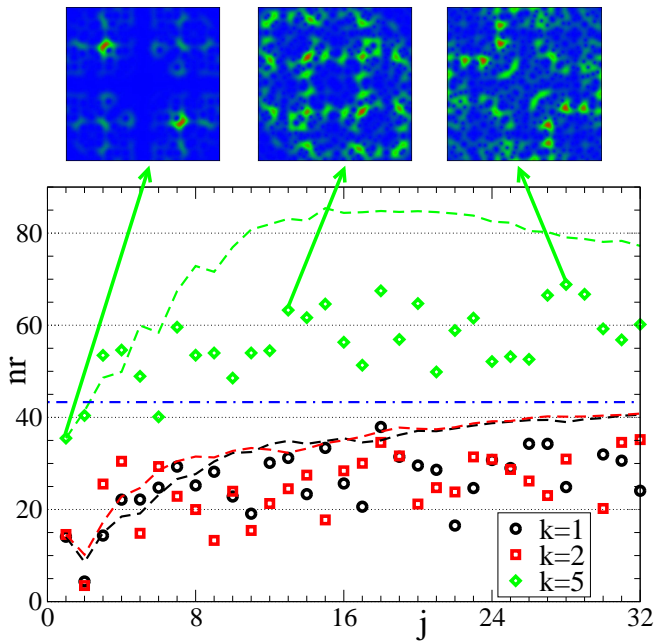} 
\end{center}
\caption{(color online) Norm ratio ($nr$) for $\hat{h}_j$ operators of 
the \emph{shift} quantum open baker maps with dimension $D=3^5=243$.
Black circles, red squares and blue diamonds represent $k=1$, $k=2$ and 
$k=5$ members respectively. $nr$ is also shown for corresponding $\hat{Q}_j$ in 
dashed lines.  
Three Husimi functions ($h_j(q,p)\equiv\vert\langle q,p\vert\hat{h}_j\vert q,p\rangle \vert$) 
for $k=5$ and $j=1$, $j=13$ and $j=28$ 
are shown going form zero (blue) to maximum (red) on top panels (from left to right). 
Horizontal dot-dashed blue line is the norm ratio value obtained for the uniform quantum distribution 
on the repeller $I_{rep}$.
}
\label{fig4}
\end{figure}

\begin{figure}%[htp]
\begin{center}
\includegraphics[width=0.48\textwidth]{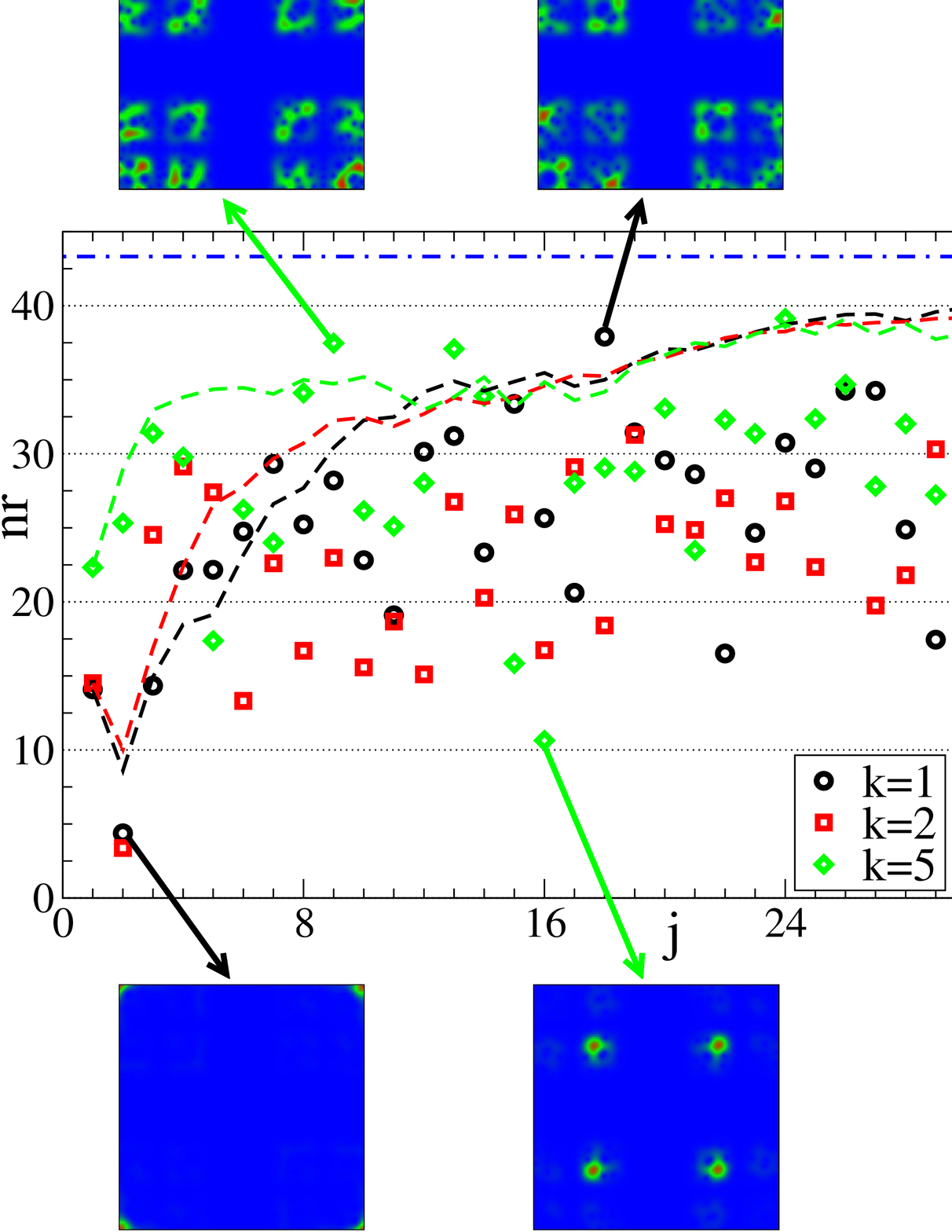} 
\end{center}
\caption{(color online) Norm ratio ($nr$) for $\hat{h}_j$ operators of 
the \emph{intersection} quantum open baker maps with dimension $D=3^5=243$.
Black circles, red squares and blue diamonds represent $k=1$, $k=2$ and 
$k=5$ members respectively. $nr$ is also shown for corresponding $\hat{Q}_j$ in 
dashed lines.  
Four Husimi functions ($h_j(q,p)\equiv\vert\langle q,p\vert\hat{h}_j\vert q,p\rangle \vert$) 
for $k=1$ and $k=5$
are shown going form zero (blue) to maximum (red). 
The most localized cases corresponding to one periodic orbit scar ($k=1$ and $j=2$) 
and two periodic orbit scar
($k=5$ and $j=16$) are shown on the right and left bottom panels respectively.
Some cases of uniformly distributed functions on the repeller, for $k=5$ and $j=9$ and 
$k=1$ and $j=18$, are illustrated on
left and right top panels respectively.
Horizontal dot-dashed blue line is the norm ratio value obtained for the uniform quantum 
distribution on the repeller $I_{rep}$.
}
\label{fig5}
\end{figure}

The phase space representation 
of both $\hat{h}$ and $\hat{Q}$ operators can be defined by means of 
coherent states $\vert q,p\rangle$ in the same way as in \cite{ermannprl}
\begin{eqnarray}
 h_j(q,p)&=&\vert\langle q,p\vert \hat{h}_j\vert q,p\rangle\vert\\
 Q_j(q,p)&=& \vert\langle q,p\vert \hat{Q}_j\vert q,p\rangle\vert
\end{eqnarray}
Clearly, the $h_j$ and $Q_j$ distributions tend to be localized on a discrete time version of the repeller 
(for Ehrenfest time $l$). Individual $h_j$ are distributed in the discrete 
time repeller in the same way as eigenstates of closed systems 
spread on the torus. Worth mentioning, scarring phenomena is visible for some $h_j$ on periodic orbits that 
belong to the repeller (see panels in Figs. \ref{fig4} and \ref{fig5}).

\begin{figure}%[htp]
\begin{center}
\includegraphics[width=0.48\textwidth]{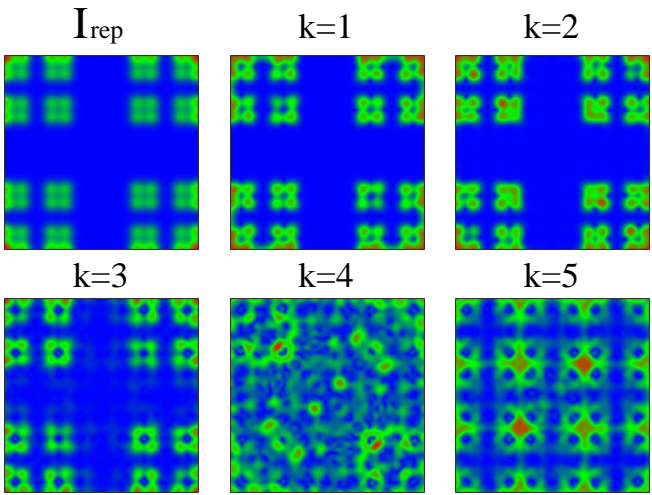} 
\end{center}
\caption{(color online) Phase space representation of $\langle q,p\vert I_{rep}\vert q,p\rangle$ (left top panel) and $Q_{32}$ operators for 
 all members of the \emph{shift} family with $l=5$. $Q$ functions are shown for $k=1,2,3,4,5$ and $j=2^5=32$ in 
a $D$-dimensional Hilbert space with $D=3^5=243$. The values of the functions goes from zero in blue to maximum value in red.
}
\label{fig6}
\end{figure}

In order to obtain a better characterization of the eigenstates we quantify the localization 
in phase space by defining the \emph{norm ratio} ($nr$) of real and positive distributions $h_j$ (and $Q_j$):
\begin{equation}
 nr(h)=\left( \frac{{\|h\|}_1/{\|h\|}_2}{{\|\rho_c\|}_1 / {\|\rho_c\|}_2} \right)^2
\end{equation}
where a coherent state is used for normalization $\rho_c=\vert q,p\rangle\langle q,p\vert$, 
and the phase space norm is defined as
\begin{equation}
 \|h\|_\gamma=\left( \int_{\mathbb{T}^2} h(q,p)^\gamma dqdp \right)^{\frac{1}{\gamma}}
\end{equation}
The norm ratio is independent of the $h$ normalization, and the position in phase space 
$(q,p)$ of the coherent state in $\rho_c$. 
Note that norm ratio is 1 for a coherent state by definition, and reaches the maximum value $D/2$ for a uniform distribution 
in a $D$-dimensional Hilbert space. For comparison purposes we define a uniform quantum distribution 
on the discrete version of the repeller as 
\begin{equation}\label{eq.irep}
 I_{rep}=\sum_{j,j^\prime}^{D}\chi_l(q_j,p_{j^\prime})\vert q_j,p_{j^\prime}\rangle\langle q_j,p_{j^\prime}\vert
\end{equation}
with $j,j^\prime=1,\ldots,D$, and where $\chi_l(q,p)$ is the finite time repeller
defined as
\begin{equation}
\chi_l(q,p)=\left\{ \begin{tabular}{cc} 1&\ \ \ \ \text{if} $(q,p)\in\mathcal{K}_l$ \\ 0&\text{otherwise} \end{tabular}\right.
\end{equation}
with discrete time version of the repeller $\mathcal{K}_l$ defined as the set of points that remain in the allowed 
region for the first $l$ iterations (both for positive and negative times).
The $I_{rep}$ operator can be interpreted as the coherent state quantization of the finite time repeller (see \cite{vallejo}).
Phase space representation for this operator, $\langle q,p\vert I_{rep}\vert q,p\rangle$, is shown in top left panel 
of Fig.\ref{fig6} for $D=3^5=243$.

We have found that for the \emph{shift} family the eigenstates corresponding to members having $k \geq l/2$ 
(i.e., for which the openings have widths smaller or equal than $\sqrt{\hbar}$) suddenly become 
strongly delocalized. By strong delocalization we mean that they show $nr$ values higher than the 
one corresponding to the $I_{rep}$ distribution of Eq. \ref{eq.irep}. That is, the 
eigenstates are not only delocalized inside the discrete classical repeller but they have finite 
probability in classically forbidden regions in phase space. This can be clearly seen in Fig. \ref{fig4}, 
where the values of $nr$ are shown for the shift family members $k=1,2,3$, for $D=3^5=243$. 
The upper panels illustrate this remarkable behavior by means of phase space portraits (Husimi functions) of selected resonances.   
We should notice that for $D=3^5$ the families at $k=5$ imply openings of ``pixel'' size and thus generate strong diffraction.
The $Q$ distribution goes asymptotically 
towards the uniform distribution on the discrete version of the repeller for $k<l/2$, but clearly changes its 
behavior above this critical value. In this latter case the finite probability on forbidden regions of phase space is 
reflected by the $nr$ values of $Q$ that saturate at about the double of that corresponding to the uniform 
quantum distribution on the discrete version of the repeller. Fig.\ref{fig6} shows the phase space respresentation of $I_{rep}$ 
and $Q_{32}$ for all members of \emph{shift} family ($k=1,2,3,4,5$).

On the other hand, the \emph{intersection} family also shows an increase in $nr$ values for $k\geq l/2$ 
but this is now a very subtle effect (though a jump in $nr$ can be seen for $Q$). All of its members have resonances distributed outside of the forbidden 
regions in phase space. They become more delocalized as $k$ increases but the $nr$ values remain always 
below the one corresponding to the uniform distribution on the discrete repeller. This is illustrated 
in Fig. \ref{fig5} and its panels that show phase space plots (Husimi functions) of selected resonances.

\begin{figure}%[htp]
\begin{center}
\includegraphics[width=0.48\textwidth]{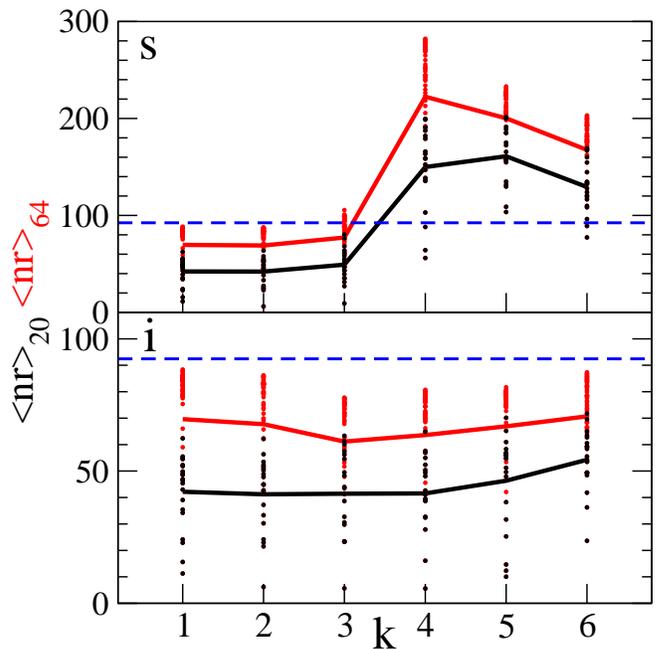} 
\end{center}
\caption{(color online) Average norm ratio $<nr>$, for both families of quantum open tri-baker maps for $D=3^6=729$.
Top and bottom panels show \emph{shift} and \emph{intersection} families respectively.
The average norm ratio for the first $20$ and $64$ eigenstates are shown in red and black lines respectively, 
while circles represents individual values of norm ratios.
Horizontal dot-dashed blue line is the norm ratio value obtained for the uniform quantum distribution on the 
repeller $I_{rep}$.
}
\label{fig7}
\end{figure}

All this is summarized in Fig. \ref{fig7}, where the average norm ratios together with 
individual values for the resonances of both families are shown (in this case we have taken 
$D=3^6=729$). It is clear that a quantum phase transition takes place at the level of the resonances 
of the \emph{shift} family. The critical value of the parameter $k$ is $l/2$, at which they 
become strongly delocalized and remain so for all values above it (we have confirm this result for $l\leq9$). 
This applies to the 
long lived portion of the spectrum as evidenced by the two limits taken in the corresponding 
average (i.e. for the first $20$ and $64$ eigenstates). 

For the case of a given \emph{shift} family member (fixed $k> l/2$), 
the localization in the forbidden area disappears in the semiclassical limit (increasing $l$), 
since $k$ will be much smaller than the new $l$ value.  
In the case of the \emph{intersection} 
family there is no such phase transition but the delocalization of the resonances slightly grows 
for $k\geq l/2$. 

The main difference that exists for \emph{shift} and \emph{intersection} families is that for one map iteration,
the allowed regions are distributed in the whole torus for the first case, 
and only lay on the discrete time repeller for the second one (see Fig.\ref{fig6}). 
In the quantum case, for openings comparable with $\hbar$, the diffraction effects are very strong
leading to notably different behaviors between both families.

\section{Conclusions}
\label{sec:Conclusions}

We have investigated the behavior of the spectrum and the eigenstates corresponding to 
classical maps that share the same classical repeller (and other asymptotic properties), 
but differ in their short time behavior. Two families of 
these maps that we call \emph{shift} and \emph{intersection} play a fundamental role 
in understanding their dependence on the shape and the area of the escape region, respectively. 
We have found that there is a quantum phase transition in the resonances of the \emph{shift} 
family. For openings having areas smaller than $\hbar$ they become strongly delocalized
in phase space, occupying forbidden regions. 
This makes difficult the quantum to classical link, even at very short times. Correspondingly, 
the spectrum oscillates in a non trivial way. The nature of the resonances for $k\geq l/2$ is 
different from that corresponding to lower values of $k$, as can be seen from Fig. \ref{fig7} (upper panel). 
This suggests that the oscillation in the spectrum does not reflect an equivalence of the 
long lived sector of the extreme cases $k=1$ and $l$. This will be further investigated 
in \cite{future}. In the case of the \emph{intersection} family we have found no such phase transition. However, there 
is a growing delocalization of the states as $k$ grows and for $k\geq l/2$ we have 
also identified a quick saturation of $nr$ values for both the eigenstates themselves and the $Q$ 
distributions towards the uniform distribution on the discrete repeller. In this family, the effect 
of the intersection with wider openings erases the probability that escapes towards forbidden regions in 
phase space and this prevents the phase transition to occur here. The spectrum 
shrinks in this case, reflecting the sudden loss of probability due to a larger area of the 
opening as $k$ grows. 

All this has very important consequences for the theory of open systems. Firstly, we mention 
that the fractal Weyl law should take into account short time dynamics for any given dimension 
$D$, perhaps in its prefactor, and not only the invariants as the 
classical fractal dimension or escape rate. Also, the 
sensitivity of the spectrum with respect to which qubit is opened could have relevant implications 
to quantum computation. On the other hand, the strongly delocalized cases of the \emph{shift} family 
pose an important question in order to understand 
the validity and performance of the semiclassical theory attempting to reproduce the spectrum 
and eigenstates of open maps. This is because this theory is based on highly localized functions, 
i.e. the so called scar functions which are constructed with the classical information on and around 
the periodic orbits that live on the repeller. This lack of localization is probably due 
to the fact that for these extreme cases any quantum manifestation of the 
stable and unstable manifolds associated to the periodic orbits is lost. 
Then, it is crucial to understand how this phenomenon manifest itself even when the 
openings are not so extremely quantum in nature. This would not only give a better idea of 
how the semiclassical approximation works, but would tell us its validity range and perhaps 
how we can improve it. Finally, open billiards constitute a very interesting field of application 
and verification of our findings \cite{openbilliards1,openbilliards2}. 
Resonant micro-cavities for lasers can be adapted in order to realize the purely quantum 
repellers found for $k\geq l/2$ \cite{future}.

%%%%%%%%%%%%%%%%%%%%%%%%%%%%%%%%%%%%%%%%%%%%%%%%%%%%%%%%%%%%%%%%%%%
\begin{acknowledgments}

Partial support by ANPCyT and CONICET is gratefully acknowledged.
\end{acknowledgments}
%%%%%%%%%%%%%%%%%%%%%%%%%%%%%%%%%%%%%%%%%%%%%%%%%%%%%%%%%%%%%%%%%%%

\end{document}